\documentclass{elsart}
\usepackage{color}
\usepackage{graphicx}
\usepackage{amsmath,amssymb}

\begin{document}

\begin{frontmatter}
\title{Net-charge probability
 distributions in heavy ion  collisions
at chemical freeze-out}
\author[gsi,emmi,tud,fias]{P.\ Braun-Munzinger},
\author[gsi]{B.\ Friman},
\author[bnl,biel]{F.\  Karsch},
\author[wroclaw,gsi,emmi]{K.\ Redlich},
and
\author[bnl]{V.\ Skokov}
\address[gsi]{GSI Helmholtzzentrum f\"ur Schwerionenforschung,
  D-64291 Darmstadt, Germany}
\address[emmi]{ExtreMe Matter Institute EMMI, GSI, D-64291 Darmstadt,
  Germany}
\address[tud]{Technical University Darmstadt, D-64289 Darmstadt,
  Germany}
\address[fias]{Frankfurt Institute for Advanced Studies, J.W.\ Goethe
  University, D-60438 Frankfurt, Germany}
\address[bnl]{Physics Dept., Brookhaven National Laboratory Upton,
  NY-11973, USA}
\address[biel]{Fakult\"at f\"ur Physik, Universit\"at Bielefeld,
 D-33501 Bielefeld, Germany}
\address[wroclaw]{Institute for Theoretical Physics, University of
  Wroclaw, 50-204 Wroclaw, Poland}

\begin{abstract}
We explore net charge probability distributions in heavy ion
collisions within the hadron resonance gas model. The distributions
for strangeness, electric charge and baryon number are derived.  We
show that, within this model, net charge probability distributions and
the resulting fluctuations can be computed directly from the measured
yields of charged and multi-charged hadrons.  The influence of
multi-charged particles and quantum statistics on the shape of the
distribution is examined.  We discuss the properties of the net proton
distribution along the chemical freeze-out line. The model results
presented here can be compared with data at RHIC energies and at the
LHC to possibly search for the relation between chemical freeze-out
and QCD cross-over lines in heavy ion collisions.
\end{abstract}


\end{frontmatter}

\section{Introduction}
\label{sec:int} 
Detailed studies of hadron production in nucleus-nucleus collisions
from SIS up to LHC energies provide the very interesting result that
particle yields exhibit thermal characteristics and are well described
by the statistical operator of a hadron resonance gas (HRG)
\cite{hwa,cleymans,star3}. Furthermore, there is a unique relation
between the collision energy and the corresponding thermal parameters,
which defines the so called chemical freeze-out line in the
temperature and chemical potential
($T,\mu$)--plane~\cite{hwa,cleymans,1gev,p1gev}. The phenomenological
freeze-out curve, along which particle yields attain their measured
values, is considered to be indicative for the QCD phase boundary.

Moreover,
results of lattice QCD studies  on the critical temperature $T_c$ at
vanishing chemical potential \cite{fodor,hotQCD}
and on the slope
parameter characterizing the dependence of $T_c$ on $\mu$
\cite{karsch2,sw}
show that, at high energies, the chemical freeze-out
curve is very close to the QCD (cross-over) phase transition. Although the
origin of this fact is not fully understood, the observation is
supported by dynamical arguments~\cite{wetterich} linking the phase
transition to  chemical freeze-out. The HRG partition
function also describes the thermodynamics of a strongly interacting
medium in the hadronic phase as obtained in lattice QCD
\cite{fk1,tawfik1,tawfik2,fk2,ratti,peter,muller}.

The proximity of the freeze-out curve to the phase boundary
suggests
that the QCD phase transition and its related
critical properties should be observable in heavy ion collisions.
However, an experimental verification of a phase change in a medium
created in such collisions requires sensitive probes. In this context,
a particular role is attributed to fluctuations of conserved charges
\cite{fk1,st,hatta,koch,asakawa,spinodal,karsch3,karsch31,new}.

Recently, it was argued  that, at high energies, the history of the
system, in particular the path through the QCD cross-over transition
from the deconfined and chirally symmetric phase to the hadronic one,
may be reflected in fluctuations of conserved charges, specifically in
their higher cumulants~\cite{karsch3,karsch31,new,Skokov:2011rq}. The
characteristic
signature of such transition  may thus be manifested in 
deviations of higher cumulants of the charge distributions from the HRG
results, if the freeze-out happens near or at the QCD phase
boundary. At vanishing chemical potential, the sixth and higher order
cumulants can be negative, also in the hadronic phase, in contrast to
predictions from HRG model, where only positive yields are
obtained~\cite{karsch3}.
It
is therefore useful to consider the HRG model results on moments of
charge fluctuations as a theoretical baseline; any deviation from this
could  be an indication for critical phenomena at the time of
hadronization~\cite{fk1,karsch3,karsch31}.

First data on cumulants of the net proton multiplicity were recently
obtained by the STAR Collaboration in Au-Au collisions at several
energies and centralities \cite{star1,star2}.  The basic properties of
the measured fluctuations, in particular of their ratios, are
consistent with HRG model expectations \cite{karsch3,star2}.

In practice, the cumulants of the net proton fluctuations are obtained
from the corresponding probability distribution \cite{r1,r2,r3,r4}.
As recently noted~\cite{ourn}, a direct comparison of the net proton
multiplicity distribution, measured in heavy ion collisions, with that
of the hadron resonance gas may provide a deeper insight into the
relation between chemical freeze-out and the QCD cross-over
transition. The comparison is facilitated by the direct link between
the net proton probability distribution and the mean number of protons
and antiprotons in the HRG model~\cite{ourn}.  This eliminates
uncertainties connected with the extraction of freeze-out conditions
and thus provides a less ambiguous method for confronting the model
results with data.

A freeze-out close to the phase boundary should be reflected in a
modification of the width and shape of the net proton probability
distribution \cite{ourn}.
In the crossover regime from hadronic matter to the quark gluon plasma
the width of the distribution is expected to be reduced,
compared to the HRG model predictions.
If, on the other hand, freeze-out occurs
near the chiral critical end point, the width could be
enhanced relative to that obtained in the hadron resonance
gas. 
The first STAR results for the net proton probability distribution
from Au-Au collision at $\sqrt s_{NN}=$ 200 GeV suggest that the
width of the experimental distribution may indeed be
reduced~\cite{ourn}.
In order to scrutinize this effect, a detailed
analysis of data on the net-proton probability distributions at
different beam energies is necessary.

In this paper we extend our previous results \cite{ourn} and examine
further the significance of the net charge probability distributions as a tool
for exploring critical phenomena in heavy ion experiments. In
particular, we consider the probability distributions related to the
fluctuations of strangeness and electric charge.  We show that, in the
hadron resonance gas, these distributions, as well as all corresponding
cumulants, can be directly linked to mean particle multiplicities. The
influence of multi-charged hadrons and quantum statistics on the
shape of net charge distributions is discussed. We compute the net
proton distribution along the freeze-out line and study its
systematics.

In the next section we derive probability distributions of conserved
charges in a thermodynamic system and apply the results to the hadron
resonance gas.  In Section III we discuss the properties of these
distributions in the HRG model and make predictions for the net proton
probability distribution and the corresponding fluctuations in heavy
ion collisions at RHIC energies and at the LHC.  Finally, in Section IV
we give our conclusions.


\section{Probability distribution of conserved charges}
\subsection{General case}

In statistical physics the  probability distribution $P(N)$  can be
obtained directly from the relation  between the canonical $Z(N,T,V)$
and grand canonical ${\cal Z}(\mu_q,T,V)$ partition functions.
The thermodynamics of a system with  fixed net charge $N$ is described by  ${ Z}(N,T,V)$, which is obtained
from the density operator
\begin{equation}
{ Z}(N,T,V) = {\rm Tr}_N e^{-\beta H}.
\label{eq1}
\end{equation}
Here  the trace  is constrained to configurations with a given net charge $N$.
The grand canonical partition function
\begin{equation}
{\cal Z}(\mu_q,T,V) = {\rm Tr} e^{-\beta H + \hat{\mu}N},
\label{eq2}
\end{equation}
where $\hat{\mu}=\mu_{q}/T$, is related to ${ Z}(N,T,V)$ by the cluster decomposition
\begin{equation}
{\cal Z}(\mu_q,T,V) = \sum_{N} {Z}(N,T,V) e^{\hat{\mu} N}.
\label{eq3}
\end{equation}
The physical meaning of each  term in the above  sum is the
probability of finding a configuration with net charge $N$. With the
proper normalization, the probability is~\cite{hwa,r1,r2}
\begin{equation}
P(N) = \frac{1}{{\cal Z}(\mu_q,T,V) }  { Z}(N,T,V) e^{\hat{\mu} N}.
\label{eq4}
\end{equation}
By eliminating
  ${\cal Z}(\mu_q,T,V)$ in favor of  the  thermodynamic pressure $\ln {\cal Z}=VT^{3} \hat p(T,\hat{\mu})$,
  one arrives at the probability distribution
\begin{equation}
P(N) = {Z}(N,T,V) e^{\hat{\mu} N - V T^{3}  \hat p(T, \hat{\mu}) }
\label{eq5}
\end{equation}
for finding a state with net charge $N$ in a system of volume $V$ at the temperature $T$.

The canonical partition function can be directly obtained from the grand
canonical one computed at imaginary chemical potential~\cite{hwa},
\begin{equation}
{Z}(N,T,V)={1\over {2\pi}}\int_0^{2\pi} d\theta e^{-i\theta N} {\cal Z} (i T \theta ,T,V),
\label{eq6}
\end{equation}
where the chemical potential  was  Wick rotated   by the substitution $\hat\mu\to i\theta$.
Equations (\ref{eq4}) and (\ref{eq6}) define  the probability
distribution of net charge in a thermal  system of volume $V$.

The $n$-th order
cumulant  $\chi_n$ of the net  charge $N$ is obtained by computing the
corresponding moments of $P(N)$. Alternatively, the cumulants are
given by the
generalized susceptibilities
\begin{equation}
\chi_n = \frac{\partial^n\hat p(T,\hat \mu)}{\partial \hat\mu^n}.
\label{PinvCGF}
\end{equation}
Thus, we have for the susceptibility or the second order cumulant,
\begin{equation}
\chi_{2}=\frac{1}{VT^3} \left(\langle N^{2}\rangle - \langle
N\rangle^{2} \right)=\frac{\partial^2\hat p(T,\hat \mu)}{\partial
  \hat\mu^2},
\label{chi2}
\end{equation}
where the moments of $P(N)$ are defined by
\begin{equation}
\langle N^{n}\rangle = \sum_{N} N^{n}P(N).
\label{momentsp}
\end{equation}

\begin{figure*}
\includegraphics*[width=3.9cm,height=5.5cm]{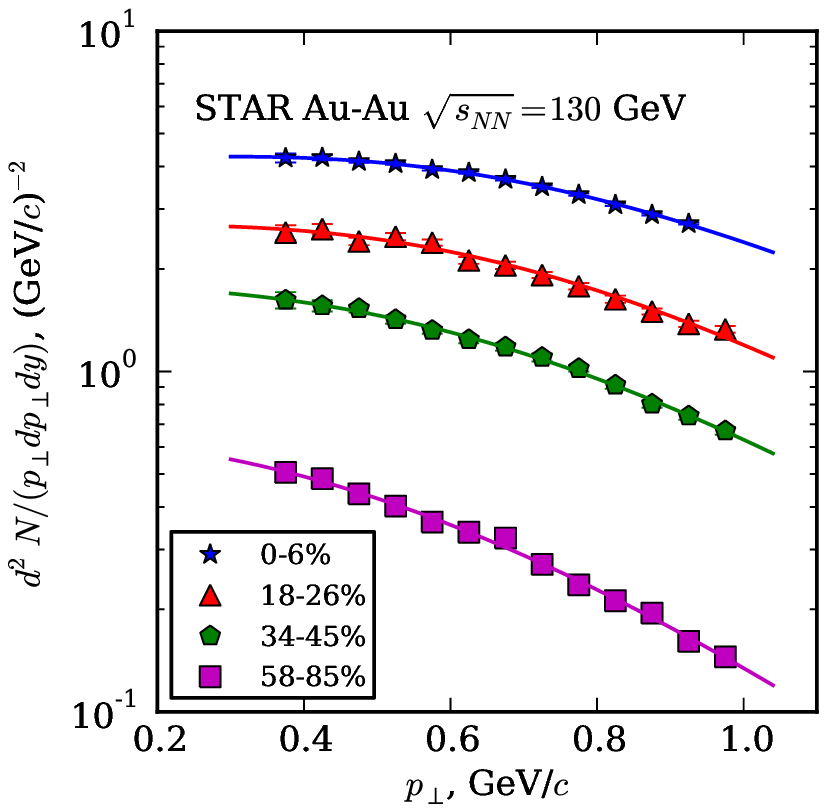}
\hspace*{.3cm}\includegraphics* [width=3.9cm,height=5.5cm]{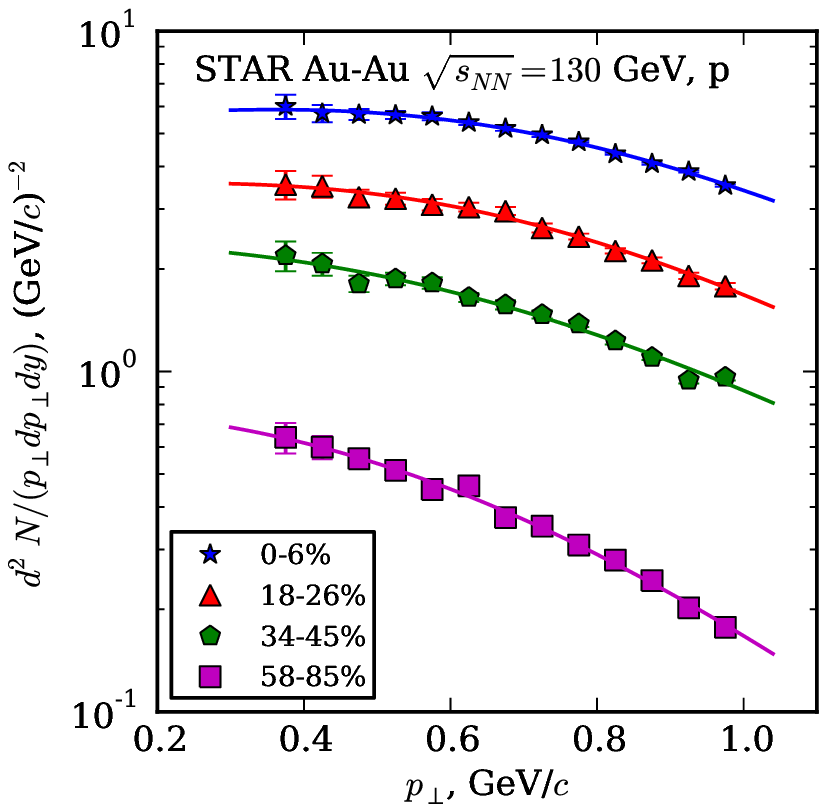}
\hspace*{.3cm}\includegraphics* [width=5.0cm,height=5.5cm]{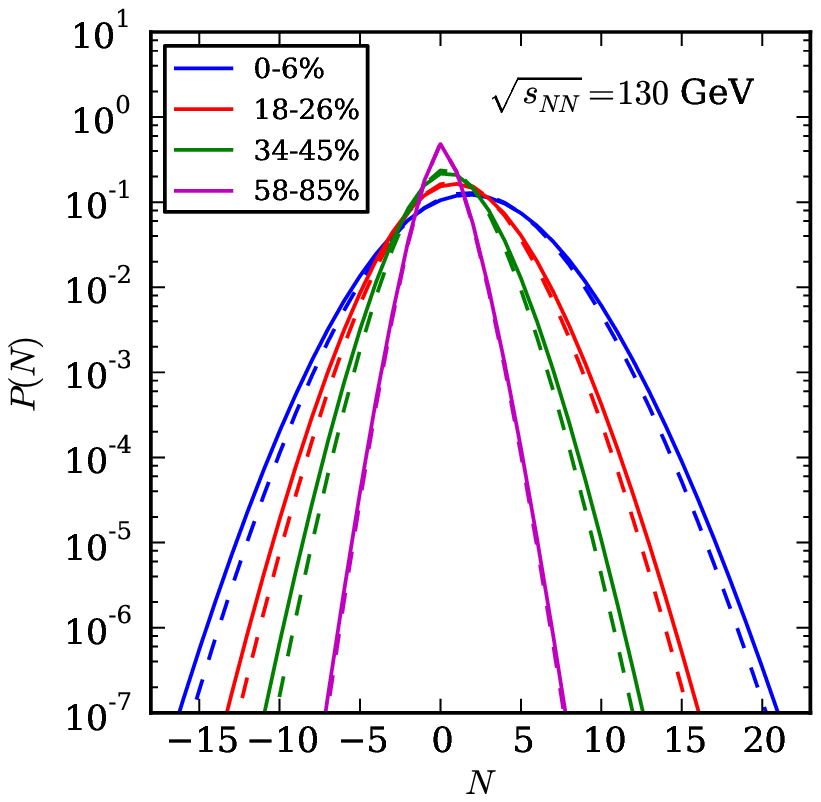}
\caption
{ Left-hand and middle figures:
  The $p_t$-distributions of  antiprotons and  protons
   in Au-Au collisions at $\sqrt s_{NN}=$130 GeV for different
  centralities \cite{star4}. The lines are blast-wave   fits \cite{blast}
 to the
  data.  Right-hand figure: The net-proton distribution in the hadron
  resonance gas computed from Eq.~(\ref{p1}) for different
  centralities.  The full lines are obtained using experimental inputs for the
  proton $\bar N_p$ and antiproton $\bar N_{\bar p}$
  yields, while the broken-lines are obtained with $ \bar N_p$ and $\bar N_{\bar
    p}$ computed in the thermal model with parameters describing the chemical
  freeze-out~\cite{star4}.  }
\label{fig130}
\end{figure*}

\subsection{The net-charge  distribution in a hadron resonance gas}

In a strongly interacting medium there are conservation laws related
to the global symmetries of the QCD Lagrangian.  In the following, we
explore probability distributions of such conserved charges and the
corresponding fluctuations. We focus mainly on fluctuations of the net
baryon number in hot and dense hadronic matter, but give results
also for fluctuations of strangeness and electric charge.

We use the hadron resonance gas partition function to model the
thermodynamics of strongly interacting matter. This model yields a
 good approximation to the lattice QCD equation of state below the
QCD phase transition.
%
%
%
%
%
In the HRG model  the
thermodynamic pressure is a sum of meson and baryon contributions. The
conservation of baryon number, strangeness and
electric charge is accounted for by introducing the corresponding chemical
potentials.



In  Boltzmann approximation, we can write the thermodynamic pressure in the following form:
\begin{equation}
 \beta Vp(T,\hat\mu)= \sum_{n=0}^{|q|}[z_n e^{n\hat\mu}+{\bar  z}_{n}\,e^{-n\hat\mu}].
\label{eq7}
\end{equation}
Here one of the conservation laws is explicitly exhibited. We denote
the corresponding chemical potential generically by $\hat \mu$.
Depending on which conservation law is picked,  baryon number,
electric charge and strangeness, the maximum charge $q$ take the
values $|q|=1, 2$ or 3, respectively.
The remaining conservation laws are implicitly included in the
parameters $z_n$ and ${\bar  z}_{n}$, which encode the thermal
phase-space
of  all particles and antiparticles carrying charge  $n$ and $-n$, respectively.
Within the HRG model, we find
 \begin{equation}
z_n= \sum_{i\in {\rm hadrons}} z_n^i,\quad  z_n^i = \frac{VT}{2\pi^2}  g_i m_i^2 K_2(m_i/T)e^{\beta\vec l_i\cdot\vec \mu_q } ,
\label{eq8}
\end{equation}
where the sum is taken over all stable hadrons as well as all hadron resonances. Moreover, the vector $\vec l_i$
represents those charges carried  by particle $i$ that are not explicitly shown in Eq. (\ref{eq7}), while $\vec \mu_q$ subsumes the corresponding chemical potentials.
The $ {\bar z}_{ n}$ is obtained from Eq. (\ref{eq8}) by the  replacement    $\vec \mu\to -\vec \mu$.

The phase-space parameters $z_n$ and ${\bar  z}_{n}$ are related to
the mean-number  of particles $\bar N_n$  and antiparticles $\bar
N_{-n}$ carrying charge $n$ and $-n$, respectively,
\begin{equation}
\bar N_n=z_{ n} \exp(n\hat{\mu}), ~~ \bar N_{-n} ={\bar z}_{n} \exp(-n\hat{\mu}).
\label{eq9}
\end{equation}

The partition function, where one conservation law is treated exactly,
is obtained using  Eqs. (\ref{eq6}) and (\ref{eq7}). We denote the
resulting partition function by $Z^{(|q|)}$, where $|q|$ is the
relevant
maximal
charge.
We start with the most complex case, i.e. with $|q|=3$.
Within the HRG model the corresponding canonical partition function is
given by~\cite{hwa}
\begin{eqnarray}
{Z}^{(3)}(N,T,V) =\sum_{i=-\infty}^\infty\sum_{k=-\infty}^\infty a_1^{-2i-3k+N} a_2^ia_3^k
I_{N-2i-3k}(x_1)I_i(x_2)I_k(x_3),
\label{eq10}
\end{eqnarray}
where $I_N(x)$ is the modified Bessel function.
 In Eq. (\ref{eq10}),  we have also introduced the notation
\begin{eqnarray}
a_n=\sqrt {{z_n}\over {{\bar  z}_{n}}}, ~~~{\rm and }~~~~x_n=2 \sqrt{z_n {\bar z}_{ n}}=2 \sqrt{\bar N_n\bar N_{-n}}
\label{eq11}
\end{eqnarray}
with $n\in (1,2,3)$.
Within the hadron resonance gas model  the canonical partition function
${Z}^{(3)}$ is appropriate  for describing a system with exact
conservation of strangeness.

For the electric charge, the appropriate canonical partition function is
${Z}^{(2)}$, since there are single- and double-charged hadrons.
{We stress, however, that this is correct only within the Boltzmann
approximation we consider here. In the electric charge sector this
approximation is not suitable, as we will discuss later.}
The generic form of $Z^{(2)}$ is obtained from Eq. (\ref{eq10}) by taking the
limit, $x_3\to 0$ and $a_3\to 1$.  This yields a
non-zero contribution only  for $k=0$, with $\lim_{x_3\to
  0}I_{k=0}(x_3)=1$, leaving the summation over $i$. The canonical
partition function for  baryon number conservation,
${Z}^{(1)}$,   is then obtained from ${Z}^{(2)}$ by taking the corresponding limit, $x_2\to 0$ and $a_2\to 1$.
Again, only the $i=0$ contribution survives, leaving an expression in closed form~\cite{r1}
\begin{equation}
\label{z1}
Z^{(1)}(N,T,V)=\left(\frac{z_{n}}{{\bar z}_{n}}\right)^{N/2}I_{N}(2\sqrt{z_{n}{\bar z}_{n}}).
\end{equation}

In the HRG  model, the probability P(N) to find a state  with  net-charge $N$
is obtained from  Eqs.(\ref{eq4}), (\ref{eq7}) and (\ref{eq10}). We compute explicitly the probability
distributions relevant for the phenomenology of heavy-ion collisions, i.e. for strangeness $(S)$, electric charge $(Q)$ and baryon number $(B)$.

Since there are hadrons with strangeness one, two and three, the relevant probability distribution is that obtained from ${Z}^{(3)}$:
\begin{eqnarray}
P(S) &=& \left({{\bar S_1}\over {\bar S_{-1}}}\right)^{ S\over 2}\exp\left[\sum_{n=1}^3\left(\bar S_n+\bar S_{-n}\right)\right]
\sum_{i=-\infty}^\infty\sum_{k=-\infty}^\infty
\left({{\bar S_3}\over {\bar S_{-3}}}\right)^{k/2} I_k\left(2\sqrt {\bar S_3\bar S_{-3}}\right)\nonumber
\\
&&\left({{\bar S_2}\over {\bar S_{-2}}}\right)^{i/2} I_i\left(2\sqrt {\bar S_2\bar S_{-2}}\right) 
%
\left({{\bar S_1}\over { \bar S_{-1}}}\right)^{-i-3k/2}I_{2i+3k-S}\left(2\sqrt{\bar S_1\bar S_{-1}}\right),
\label{p3}
\end{eqnarray}
where $\bar S_s$ and $\bar S_{-s}$ denote the mean-number of particles and antiparticles carrying strangeness $s$, where $s\in (1,2,3)$ (see    Eq. (\ref{eq9})).

For the electric charge, {treated in Boltzmann approximation,}
the corresponding probability is given by
\begin{eqnarray}
P(Q) &=& \left({{\bar Q_1}\over {\bar Q_{-1}}} \right)^{Q\over 2}\exp\left[\sum_{n=1}^2\left(\bar Q_n+\bar Q_{-n}\right)\right] 
\sum_{i=-\infty}^\infty
\left({{\bar  Q_{2}}\over {\bar Q_{-2}}}\right)^{i/2} I_i\left(2\sqrt {\bar Q_2\bar Q_{-2}}\right)\nonumber
\\
&&\left({{{\bar Q_1}}\over { \bar Q_{-1}}}\right)^{-i}I_{2i-Q}\left(2\sqrt{\bar Q_1\bar Q_{-1}}\right),
\label{p2}
\end{eqnarray}
where $\bar Q_q$ and $\bar Q_{-q}$ denote the mean number of particles and antiparticles with electric charge  $q$ and $-q$ respectively, with $q\in (1,2)$.

Finally,  the probability $P(B)$  of  net-baryon number $B$ can be expressed in terms of the mean-number of baryons $\bar B_1$ and antibaryons $\bar B_{-1}$ \cite{ourn}  by  the Skellam distribution,
\begin{eqnarray}
P(B) &=& \left({{\bar B_1}\over {\bar B_{-1}}}\right)^{B\over 2}I_{B}\left(2\sqrt{\bar B_1\bar B_{-1}}\right)
\exp\left[\bar B_1+\bar B_{-1}\right].\nonumber\\
\label{p1}
\end{eqnarray}

Equations (\ref{p3})--(\ref{p1}) determine the probability distributions for the
fluctuations of net strangeness, electric charge and  baryon number in a hadron resonance
gas in equilibrium and under  Boltzmann statistics.
We note that the
probability distributions of the HRG depend only on the mean-number of charged particles
and antiparticles.





\section{Net-proton  distribution in heavy ion collisions}
\subsection{Probability distributions}

The mean number of particles $\bar N_n$ and
antiparticles $\bar N_{-n}$ obtained in experiment can, when available, be used
directly in Eqs. (\ref{p3}--\ref{p1}), where the volume and all thermal parameters
have been eliminated. Thus, predictions of the HRG model can
be compared with data in an unambiguous way, avoiding further model
assumptions.  In particular, using measured multiplicities $\bar N_n$ and
$\bar N_{-n}$ has the advantage that experimental cuts are
included consistently for all quantities.
We note that for a comparison of data to theory the experimental
results have to be corrected for effects due to overall baryon, strangeness, and
charge conservation.

Alternatively, in the HRG model the mean number of particles $\bar N_n$ and
antiparticles $\bar N_{-n}$ entering $P(N)$ in
Eqs. (\ref{p3}--\ref{p1}) can, given the thermal and volume parameters,
be computed using Eq. (\ref{eq9}).  In applications of the HRG model
to heavy-ion phenomenology the thermal parameters are determined along
the chemical freeze-out curve. The volume can then be fixed to
reproduce the net charge number or particular particle yields.


At present, the first approach can be applied only for the net baryon
multiplicity distribution.  The electric charge and strangeness
probability distributions require experimental input on yields of all
charge states in the acceptance window where the probability
distributions are obtained.  At present, such data are not available
for strangeness and electric charge.  On the other hand, first data on
charge fluctuations and higher order cumulants of net proton
multiplicities were recently obtained by the STAR Collaboration in
Au-Au collisions at several collision energies \cite{star1,star2}.  The
data were taken at mid-rapidity in a restricted range of transverse
momenta, $0.4~{\rm GeV} \le p_T \le 0.8~{\rm GeV}$.

Consequently, in the following, we
focus on the net-proton
probability distributions in the hadron resonance gas model.

The probability distribution, Eq.~(\ref{p1}), is readily generalized to
protons in a momentum window. Due to the factorization of the
grand canonical single-particle partition function in momentum space,
Eq.~(\ref{eq3}) yields a normalized probability distribution, when
both partition functions are restricted to the same momentum
window. Similarly, the limitation to protons follows from the
factorization of the contribution of any particle species to ${\cal Z}$.
Consequently, in the HRG model the probability distribution of net proton number in a
momentum window,
$\delta N_p= \bar N_p -\bar N_{\bar p}$,
is obtained from Eq.~(\ref{p1}) by
simply replacing the mean number of baryons $\bar B_1$ and
anti-baryons $ \bar B_{-1}$ by that of protons $\bar
N_p$ and antiprotons $\bar N_{\bar p}$ in the same kinematic window.


Thus, we can compute the net-proton distribution using Eq.~(\ref{p1}),
provided we have access to the mean values $\bar N_p $ and $\bar
N_{\bar p}$ measured in the same kinematic window.  Recently, such
studies were performed in Ref.  \cite{ourn}, using the STAR data~\cite{star1,star4} on $p_t$-distributions of
antiprotons, the mean number of net protons as well as $P(N)$ for Au-Au collisions at
$\sqrt{s_{NN}}=$ 200 GeV and at several centralities.  To assess the
dependence of $P(N)$ in A-A collisions on beam energy, we extend these
studies to different energies along the chemical freeze-out curve.

Unambiguous predictions of the model for the net proton probability
distribution can be obtained in Au-Au collisions at $\sqrt s_{NN}=$
130~GeV as well as in Pb-Pb collisions at the LHC, since at these
energies the $p_t$-distributions of protons and antiprotons are
available \cite{star4,lhc}.

In Fig.~\ref{fig130} we show the $p_t$-spectra of  protons and
antiprotons at $\sqrt s_{NN}=$ 130 GeV  obtained by the STAR Collaboration
\cite{star4}.
By fitting these  spectra with the blast-wave model \cite{blast}
and integrating the data
in the range $0.4~{\rm GeV}\leq p_t\leq 0.8~ {\rm GeV}$,
we find $\bar N_{ p}$ and $\bar N_{\bar p}$
for the different centralities.

In the right panel of Fig.~\ref{fig130} we show the corresponding HRG
result for the net proton multiplicity distribution in Au+Au
collisions at $\sqrt{s_{NN}}=$130. Also shown in this figure are the
results for $P(N)$ obtained within the HRG model with $\bar N_p$ and
$\bar N_{\bar p}$ computed at the chemical freeze-out using the
thermal parameters of Ref. {\cite{star4}}. The volume was fixed by
requiring that the measured mean-number of protons be reproduced. The
overall agreement of the two methods is satisfactory. Nevertheless, a
direct calculation of $P(N)$ from the measured yields of protons and
antiprotons is preferable, since it is unaffected by systematic errors
in the thermal parameters.

\begin{figure*}
\includegraphics*[width=6.4cm,height=6.0cm]{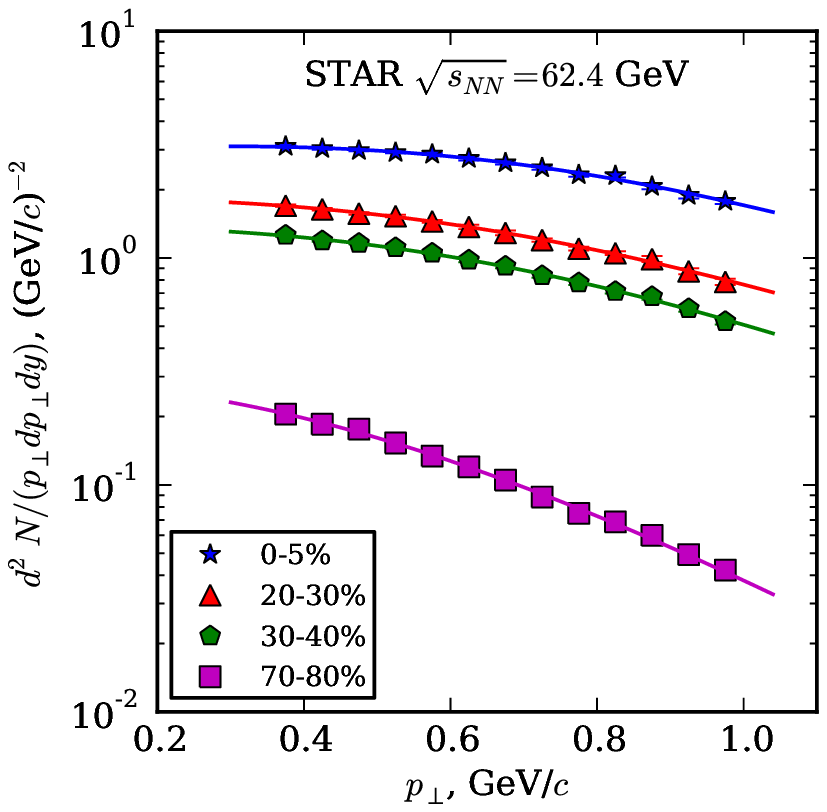}
\hspace*{0.7cm}\includegraphics* [width=6.5cm,height=6.0cm]{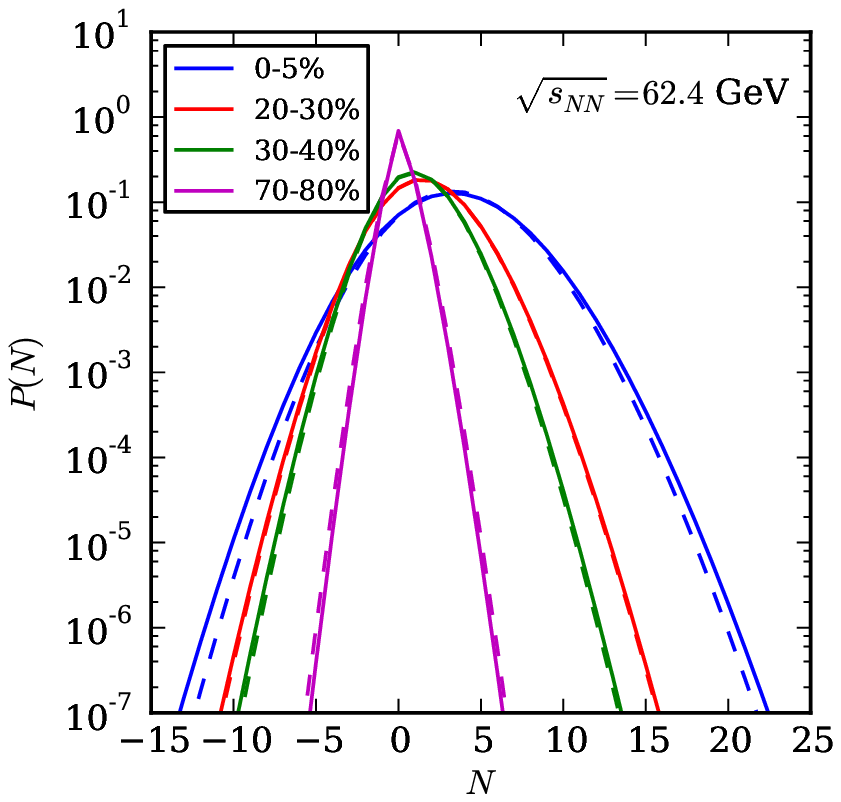}
\caption
{ Left-hand figure: The $p_t$-distribution of antiprotons obtained by the
  STAR Collaboration in Au-Au collisions at $\sqrt s_{NN}=62.4$ GeV
  for different centralities \cite{star4}. The lines are blast-wave
  fits \cite{blast} to the data.  Right-hand figure: The net-proton distribution
  calculated in the hadron resonance gas from Eq.~(\ref{p1}) for Au-Au
  collisions at $\sqrt s_{NN}=62.4$ GeV at different centralities.
  The full-lines are obtained with experimental inputs for the proton $ \bar
  N_p$ yields and the mean net baryon number $M=\bar N_p-\bar N_{\bar p}$, while the broken-lines
  are obtained with $ \bar N_p$ and $ \bar N_{\bar p}$ calculated in
  the thermal model with parameters describing the chemical freeze-out~\cite{star4}.  }
\label{fig62}
\end{figure*}


In order to find $P(N)$ in Au-Au collisions at $\sqrt s_{NN}=$ 62.4
GeV, we used the measured $p_t$-spectra of antiprotons \cite{star4} to
determine the antiproton multiplicity. The proton yield was then
fixed using the measured mean net proton number
\cite{star1,star2}. We stress that the data on net-proton number are
not corrected for efficiency \cite{star1,starn}. This implies a systematic
uncertainty in our results.  The resulting distributions for different
centralities are shown in Fig. \ref{fig62}. Again, we also show
results of the thermal model, where $\bar N_p$ and $\bar N_{\bar p}$
are computed at the chemical freeze-out using the thermal parameters of
Ref. \cite{star4}.  The two methods give similar results on $P(N)$ at
all centralities. 

Recently, the STAR Collaboration has also presented preliminary data
on moments of the net-proton distribution in central Au-Au collisions
at $\sqrt s_{NN}=$7.7, 11.5 and 39 GeV. However, at these energies the
corresponding $p_t$-spectra of protons and antiprotons are not
available. Thus, for these energies we computed $P(N)$ in the hadron
resonance gas using the proton and antiproton yields obtained in
the thermal model at chemical freeze-out with the thermal parameters
of Ref. \cite{cleymans}.
The volume was fixed by fitting the mean
number of net protons to the STAR data~\cite{bed}.  The resulting
probability distributions $P(N)$ for these three energies are shown in
Fig. \ref{all}.

In order to assess the  possible role of  dynamical effects on the
probability distribution, we compare in Fig. {\ref{urqmd} the STAR data  { on $P(N)$ in Au-Au collisions at
$\sqrt s_{NN}=$200 GeV}
and the HRG
results of Ref. \cite{ourn} with the net proton probability
distribution obtained with the UrQMD event generator.  Deviations of
the thermal model from the data and their possible origin were  discussed
in Ref. \cite{ourn}.
{ We note that the distribution obtained in the UrQMD model is much narrower than the data. The reason for this remains to be clarified.}

In Fig. \ref{all} we summarize the HRG model results for the net
proton distributions in central nucleus-nucleus at energies spanning
from $\sqrt s_{NN}=$7.7 GeV up to $\sqrt s_{NN}=$2.74 TeV.  The
probability distribution at the LHC energy is computed with proton
and antiproton yields obtained by integrating the preliminary data on $p_t$-spectra of the ALICE
Collaboration \cite{lhc} in the $p_{t}$ range
$(0.4~{\rm GeV} <p_t<0.8 ~{\rm GeV})$. The $P(N)$ distributions shown
in Figs (\ref{fig130}-\ref{all}) can be compared directly to properly
normalized data from RHIC and LHC.

\begin{figure*}
\includegraphics*[width=13.4cm,height=6.0cm]{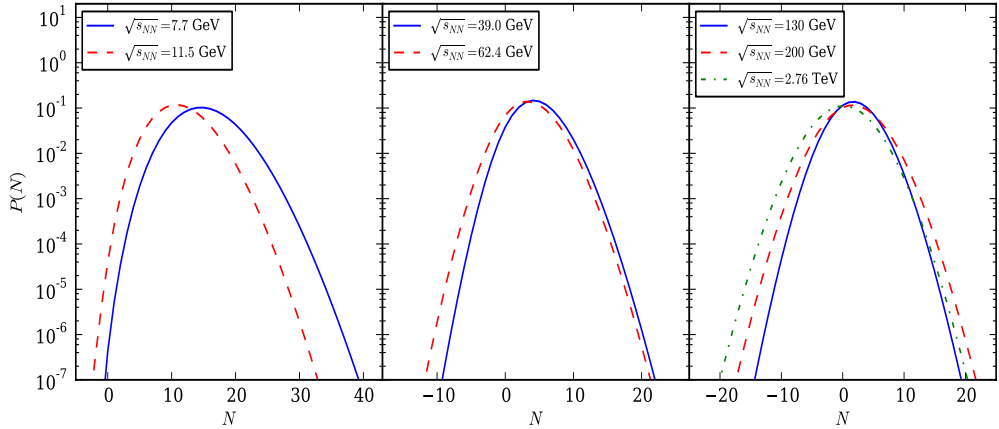}
\caption
{ Predictions of the hadron resonance gas model for the net proton
  probability distribution in central nucleus-nucleus collisions at
  different energies. The results for Au-Au collisions at $\sqrt
  s_{NN}=62.4$ and $\sqrt s_{NN}=130$ GeV are obtained using
  Eq. (\ref{p1}) with experimental inputs for the proton $\bar N_p$ and
  antiproton $\bar N_{\bar p}$ yields extracted from the $p_t$ spectra of the STAR
  Collaboration \cite{star4}.  The probability distribution for Pb-Pb collisions at
  $\sqrt s_{NN}=2.74$ TeV is obtained using Eq. (\ref{p1}) with
  experimental inputs for $\bar N_p$ and $\bar N_{\bar p}$ extracted from the preliminary
  $p_t$-spectra of the ALICE Collaboration \cite{lhc}. At lower energies,
  $P(N)$ is computed with $\bar N_p$ and $\bar N_{\bar p}$ determined at the chemical
  freeze-out with thermal parameters from Ref. \cite{cleymans}.  The
  required experimental inputs on mean net-proton numbers are from
  STAR Collaboration \cite{star1,star2}.}
\label{all}
\end{figure*}


\subsection{Beam energy systematics  of $P(N)$ and its  cumulants in heavy ion collisions }

As seen in Fig. \ref{all}, there are systematic changes in the net-proton
probability distributions in central heavy ion collisions with
increasing beam energy. There is a clear asymmetry in $P(N)$ at lower
energies, which is largely eliminated at the LHC. With increasing
$\sqrt s_{NN}$, there is a corresponding shift of the maximum of $P(N)$
from a nonzero value of $N$ towards $N=0$. These properties of $P(N)$ are direct consequences of the
decreasing net baryon density with increasing beam energy.

A rather striking feature, seen in
Fig. \ref{all}, is that
{in the entire energy range from the lowest RHIC energies up to
LHC energies}
the maximum of $P(N)$, is almost independent of energy.
This is unexpected,
{as it is well known that in heavy ion
collisions the total {\em density} of baryons plus antibaryons in the
whole $p_t$--range at mid-rapidity is approximately constant along
the chemical freeze-out curve  \cite{pbm}.}
The yield of both protons and anti-protons thus changes considerably
with energy.  {On the other hand, as we show in the following,
the maximum of $P(N)$ is proportional to $\left[
{\bar N_p+\bar N_{\bar p}} \right]^{-1/2}$}.



The maximum value of $P(N)$ can be obtained analytically from Eq. (\ref{p1}) in
two limiting cases: for very high and very low energies.  The probability
$P(N)$ has its maximum at some value of $N\approx[\langle N\rangle]$,
where $[N]$ denotes an integer part of $N$.

At high energy, the mean net-proton number is small, $\langle N\rangle=0$, while $\bar N_p$ is  large.
Therefore,  the maximum of $P(N)$ appears  at  $N\approx[\langle N\rangle]=0$ and
\begin{eqnarray}
P_{max}^{\sqrt s\to\infty} &&\simeq ~ I_0\left(2\sqrt{\bar N_p \bar N_{\bar{p}}} \right) e^{-(\bar N_p+\bar N_{\bar{p}})}
\approx {\left[2\pi (\bar N_p+\bar N_{\bar{p}})\right]^{-1/2} },
\end{eqnarray}
where we used the asymptotic expansion of the Bessel function \cite{table},
\begin{equation}
I_\nu(x\gg\nu+1) \approx \frac { e^x}{\sqrt{2 \pi x}} \; ,
\end{equation}
as well as    that  $\bar N_p\approx \bar N_{\bar{p}}$ and   $ 2
\left[\bar N_p\ \cdot  \bar N_{\bar{p}} \right]^{1/2} \approx (\bar N_p +  \bar
N_{\bar{p}})$.

At low energies, on the other hand, the  number of antiprotons is very small $\bar N_{\bar p}\ll \bar N_{{p}}$.
Consequently,  the argument of the Bessel function in Eq. (\ref{p1}) is much less than its order.  Then   \cite{table}
\begin{equation}
I_\nu(x) \approx  \frac{(x/2)^\nu} {\Gamma(\nu+1)} \approx \frac{(x/2)^\nu e^\nu} {\sqrt{2\pi\nu}\nu^\nu},
\end{equation}
and the maximum probability at low energy is obtained from
 \begin{eqnarray}
P_{max}^{\sqrt{s}\to 0} 
 \approx
\frac{ (\bar N_p \bar N_{\bar{p}})^{\bar N_p/2} e^{\bar N_p}  } {\left[{2\pi \bar N_p}\right]^{1/2} \bar N_p^{\bar N_p}} \left( \frac{\bar N_p}{\bar N_{\bar{p}}} \right)^{\bar N_p/2} e^{-\bar N_p}
\approx {\left[{2\pi (\bar N_p+\bar N_{\bar{p}})}\right]^{-1/2} },
\end{eqnarray}
where we use $\bar N_p\pm \bar N_{\bar p}\approx \bar N_p$.

Thus,  at  high as well as at low energies,  the maximal value of the
net proton  probability distribution
\begin{equation}
P_{max} \approx  \frac{1}{\left[{2\pi (\bar N_p+\bar N_{\bar{p}})}\right]^{1/2}}
\end{equation}
is approximately equal and determined by the total mean number of protons and  antiprotons.

At intermediate energies, none of the approximations used above
hold. However, near the maximum, we can approximate $P(N)$ by a
Gaussian,

\begin{equation}
P(N) \approx  \frac{1}{\sqrt{2\pi \sigma^2 }} e^{-\frac{(N-\bar N)^{2}}{2\sigma^2}},
\end{equation}
with the maximum value
\begin{equation}
P_{max} = \frac{1}{\sqrt{2\pi \sigma^2 }}.
\end{equation}
In the hadron resonance gas, the
variance is directly related to the total number of protons and antiprotons
\begin{equation}
\sigma^2=\bar N_p+\bar N_{\bar p}.
\end{equation}
Consequently, we conclude that the maximum value of the net proton probability distribution
is, to a good approximation, determined by
\begin{equation}
P_{max}^{HRG} \approx  \frac{1}{\left[{2\pi (\bar N_p+\bar N_{\bar{p}})}\right]^{1/2} }
\end{equation}
irrespective of the collision energy.
{On the other hand, as $P_{max}^{HRG}$ depends only weakly on the beam
energy (see Fig. \ref{all}), the value of $\bar N_p+\bar N_{\bar{p}}$ in the
acceptance window has to be essentially energy independent.}

In Fig. {\ref{moments}}-right we show the energy dependence of the variance $\sigma=\left[
{\bar N_p+\bar N_{\bar p}} \right]^{1/2}$ in heavy ion collisions obtained at mid rapidity with $|y|<0.5$, in the
transverse momentum window, $0.4 ~{\rm GeV}<p_t<0.8~{\rm GeV}$. The
small variation of $\sigma$, seen in this figure, is consistent with
the observed weak dependence of the maximum of $P(N)$ on $\sqrt s_{NN}$,
{which thus seems to be} a consequence of the particular
$p_t$--cut applied to the transverse-momentum distributions of protons and
antiprotons.


\begin{figure}
\begin{center}
\includegraphics*[width=8.0cm,height=6.0cm]{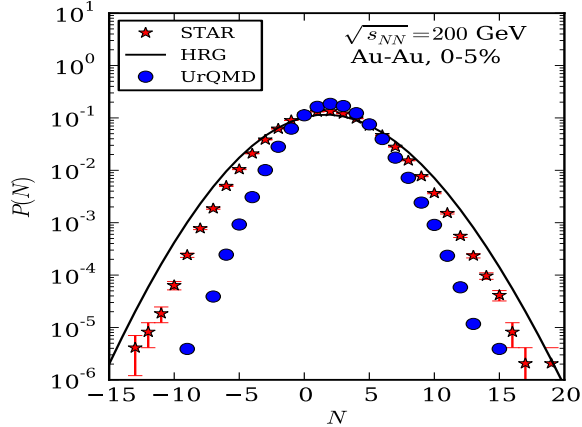}
\caption{ The net-proton probability distribution in central
  Au-Au collisions at $\sqrt s_{NN}=$ 200GeV.  The HRG results are
  from Ref. {\cite{ourn}}, while the data are due to the STAR Collaboration
  \cite{star1}. Also shown are results obtained with the UrQMD event
  generator \cite{urqmd} from a sample of $10^5$ events.  }
\label{urqmd}
\end{center}
\end{figure}


In the hadron resonance gas, not only the variance, but also all higher
moments of the net proton distribution can be expressed in terms of
the mean number of protons and antiprotons. Indeed, using
Eqs. (\ref{PinvCGF}) and (\ref{eq7}), one finds for $n=1$,

\begin{equation}
\chi_{2n}=\frac{1}{VT^3} \left( \bar N_p+\bar N_{\bar p} \right),~~\chi_{2n+1}=\frac{1}{VT^3} \left(  \bar N_p-\bar N_{\bar p}\right).\label{kmom}
\end{equation}

Consequently, the skewness $S= VT^3 \chi_{3}/( VT^3 \chi_{2})^{3/2}$ and the kurtosis $\kappa=VT^3\chi_{4}/(VT^3\chi_{2})^{2}$
are directly related to $\bar N_p$ and $\bar N_{\bar p}$ by
\begin{equation}
S=\frac{\bar N_p-\bar N_{\bar p}}{(\bar N_p+\bar N_{\bar p})^{3/2}}, ~~ \kappa=\frac{1}{\bar N_p+\bar N_{\bar p}}.\label{kur}
\end{equation}

The energy dependence of skewness and kurtosis is shown in
Fig. {\ref{moments}}-right.  The weak energy dependence of $(\bar N_p+\bar
N_{\bar p})$ implies that the kurtosis and all even cumulants are
essentially independent of beam energy. By contrast, the skewness and
all odd cumulants are strongly decreasing functions of energy, and
approach zero at high energies, where $N_{\bar p}\simeq N_{p}$.

We have seen that in the HRG model all moments of net proton
distribution can be expressed directly in terms of the mean-number of
protons and antiprotons. This can be generalized to moments of the
fluctuations of strangeness and electric charge. Indeed, using
Eqs. (\ref{PinvCGF}) and (\ref{eq7}) one finds, that in general
\begin{equation}
\chi_{2k}=\frac{1}{VT^3}\sum_{n=1}^{|q|}(\bar N_n+\bar N_{\bar n})\,n^{2k},~~\chi_{2k+1}=\frac{1}{VT^3}\sum_{n=1}^{|q|}(\bar N_n-\bar N_{\bar n})\,n^{2k+1}.\label{mg}
\end{equation}
where $|q|=1,2$ or 3 for baryon number, electric charge and strangeness, respectively.
For the electric charge fluctuations this implies that
\begin{eqnarray}
\chi_1^{Q} &=& \frac{ \bar N_1-\bar N_{-1} + 2 (\bar  N_2-\bar N_{-2})  }{VT^3}, ~~ \chi_2^{Q} = \frac{ \bar  N_1+\bar N_{-1} + 4 ( \bar  N_2+\bar N_{-2})  }{VT^3} \label{first} \\
 \chi_3^{Q} &=& \frac{ \bar N_1-\bar N_{-1} + 8 ( \bar N_2-\bar N_{-2})   }{VT^3} ,~~
 \chi_4^{Q} = \frac{ \bar N_1+\bar N_{-1} + 16 ( \bar N_2+\bar N_{-2})  }{VT^3} ,\label{last}
\end{eqnarray}
where $\bar N_{1}$ ($\bar N_{-1}$) and $\bar N_{2}$ ($\bar N_{-2}$) are the mean numbers of particles with charge $+1(-1)$ and   $+2(-2)$ respectively.
Thus, in the HRG model, the  experimental input on the  mean number of charged and multi-charged particles
is sufficient to compute the net charge probability distribution  and the corresponding moments. {We stress, however, that this is only the case in the
Boltzmann approximation, which in the case of electric charge is a
poor approximation.}


\begin{figure}
\begin{center}
\includegraphics*[width=6.0cm]{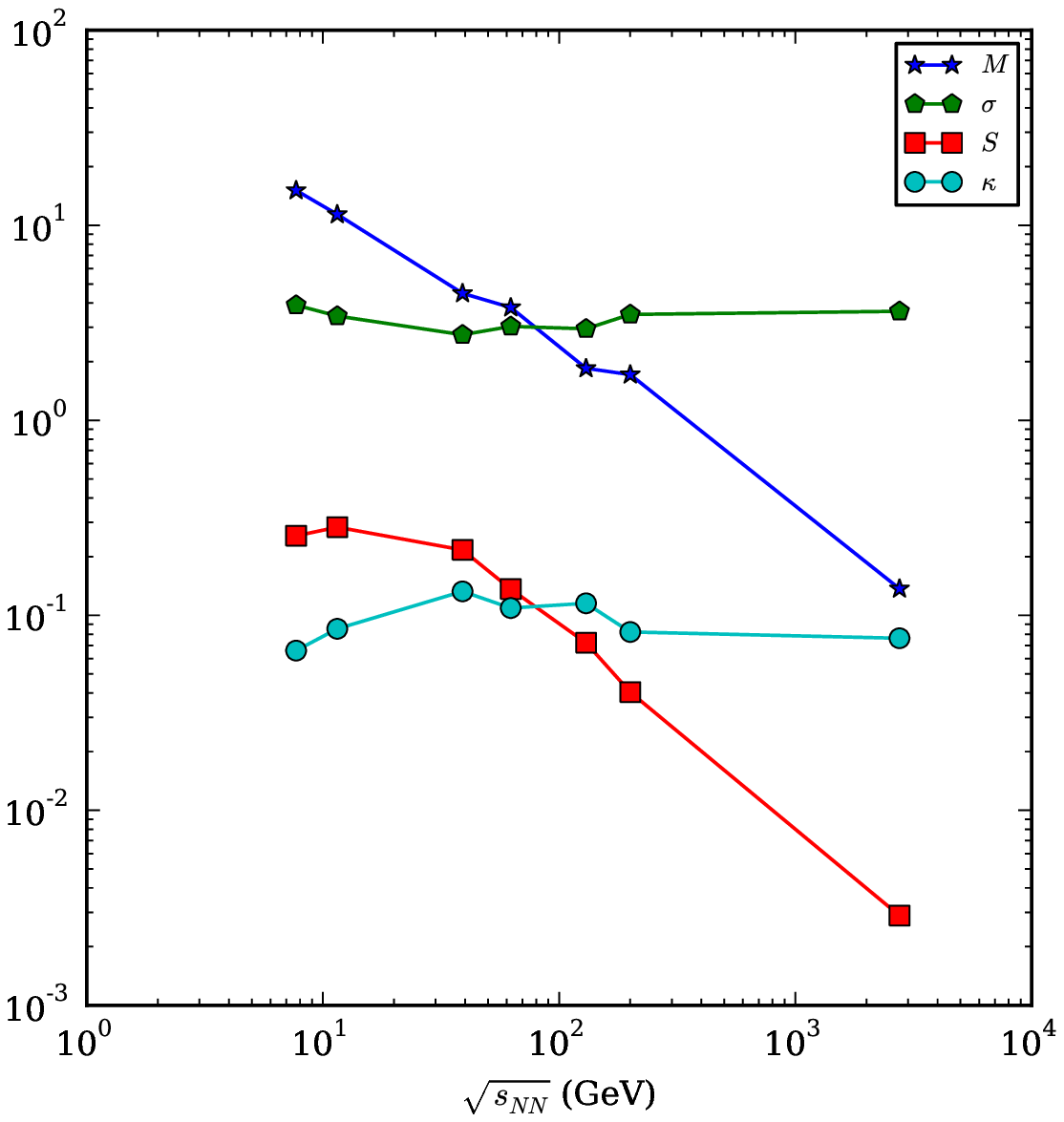}\hspace*{0.5cm}
\includegraphics*[width=6.7cm,height=6.5cm]{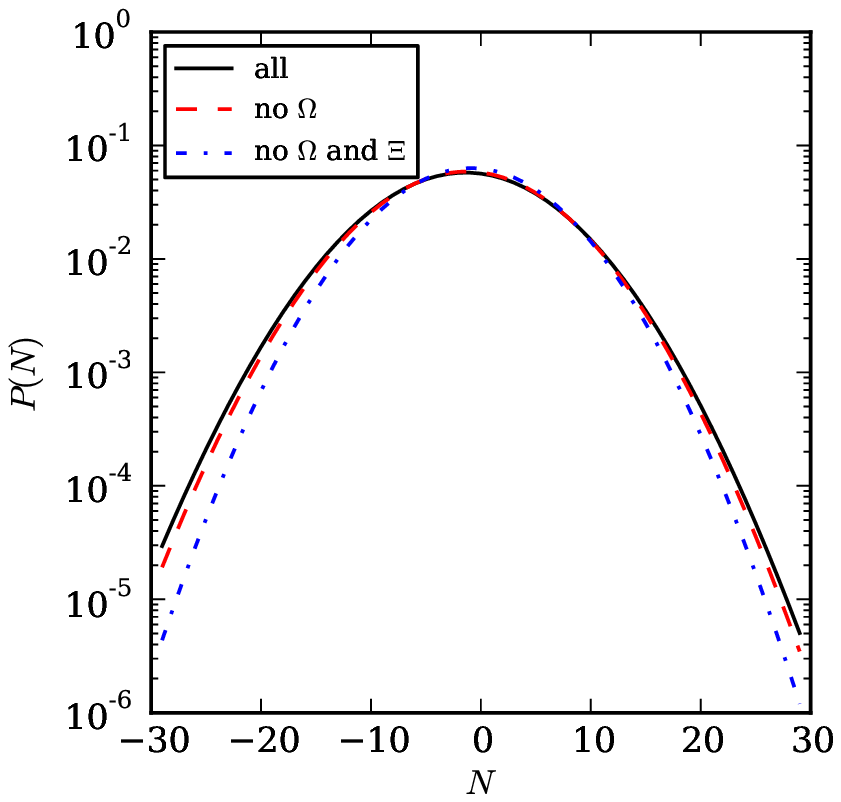}
\caption{
Left-hand figure: The mean (M), variance $(\sigma)$, skewness (S) and kurtosis
$(\kappa)$ {for net proton number}
in central Au-Au collisions computed using the probability
distributions shown in Fig.  \ref{all}. Right-hand figure: The probability distribution for net-strangeness  in a gas composed of
$(K^+,\Lambda,\Xi,\Omega)$ and their antiparticles at $T\simeq m_\pi$.
The  distribution is calculated with and  without multi-strange baryons.
}
\label{moments}
\end{center}
\end{figure}

\section{The influence of multi-charged particles and quantum statistics on $P(N)$}
The influence of multi-charged particles on fluctuations was discussed
in Ref. \cite{fk1}. In the HRG model, the product of kurtosis and variance of the net baryon
number, $\kappa\sigma^{2}$, which receives contributions from singly charged particles
only, is always equal to unity. This result does not depend on the hadron mass spectrum
nor on the thermal parameters. However, for fluctuations of the
electric charge, the contribution of particles with charge two,
$\Delta^{++}$ and $\bar\Delta^{--}$, leads to a non-trivial $T$ and
$\mu$ dependence of this quantity \cite{fk1}.  Furthermore, a
straightforward calculation shows that in an ideal pion gas, $\kappa \sigma^{2}$
is temperature dependent and differs from unity, due to quantum
statistics effects.

Clearly these effects should be reflected in the corresponding probability
distribution. To illustrate the influence of multi-charged particles
and of quantum statistics we consider two simple models: {\em i)} a
gas composed of $\Lambda, \Xi$ and $\Omega$ and their antiparticles
and {\em ii)} an ideal pion gas. Within the first model we compute
$P(S)$ and within the second $P(Q)$.
In Figs. \ref{moments}-right and \ref{fig_rat}-right we show the probability
distributions of the net strangeness and net electric charge calculated in
the models defined above.

To assess the influence of multi-charged particles, we compare the
results with $P(N)$ computed using Eq. (\ref{p1}), which accounts
only for singly charged particles.  Since the multi-charged particles
are heavy, their contribution to the mean-charge is
negligible. Consequently, the position of the maximum of $P(S)$ is not
changed when these states are included. However, the contribution of
double- and triple-charged particles modifies the tail of the
probability distribution considerably, and therefore also the higher
order moments.  Thus, a comparison of fluctuations in the hadron
resonance gas model with data, must account for multi-charged
particles, since they are included in the data through their decay
products.

The probability distributions in Eqs.  (\ref{p3}--\ref{p1}) are valid
only in  Boltzmann approximation.  This approximation is justified
for the net strange and net baryon number probability
distributions. However, owing to the small mass of the pion, quantum
statistics must be included in studies of charge fluctuations.  The
effect of the corrections to the Boltzmann approximation due to
quantum statistics is similar to that induced by multi-charged states, and
could affect the net charge distribution $P(Q)$.

 { Indeed,  the  pressure of charged pions  at temperature $T$ and electric charge
 chemical potential $\mu$ can be expanded in powers of the fugacity, $e^{\pm\mu/T}$,
\begin{equation}
\frac{p_{\pi^+} +  p_{\pi^-}}{T^4} = \frac{1} {\pi^2}
\sum_{l=1}^{\infty} d_l  (\beta m_l)^2  K_2 (\beta m_l) \cosh ( q_l \beta
\mu),
\label{ppion_recast}
\end{equation}
where $m_l = l \,m_\pi$,  $d_l=1/l^4$  and  $q_l=l$.
Thus, the pion pressure,  Eq.~(\ref{ppion_recast}), can be viewed as a sum of Boltzmann
contributions owing to particles with charge $q_{l}$
mass $m_{l}$ and  degeneracy factor $d_{l}$.

\begin{figure*}
\begin{center}
\includegraphics*[width=5.5cm,height=5.6cm]{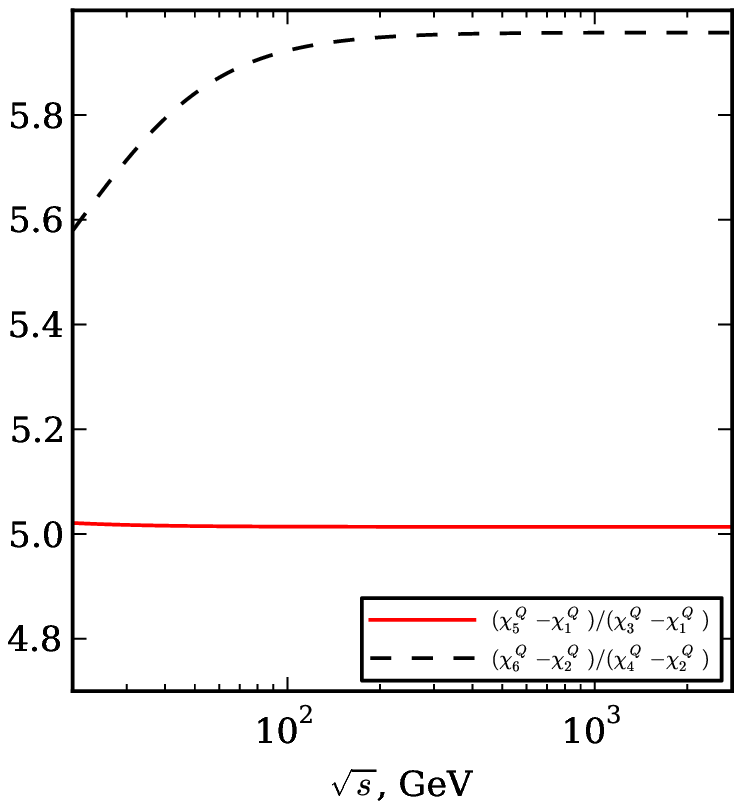}
\hspace*{ 0.6cm}
\includegraphics*[width=6.2cm,height=5.7cm]{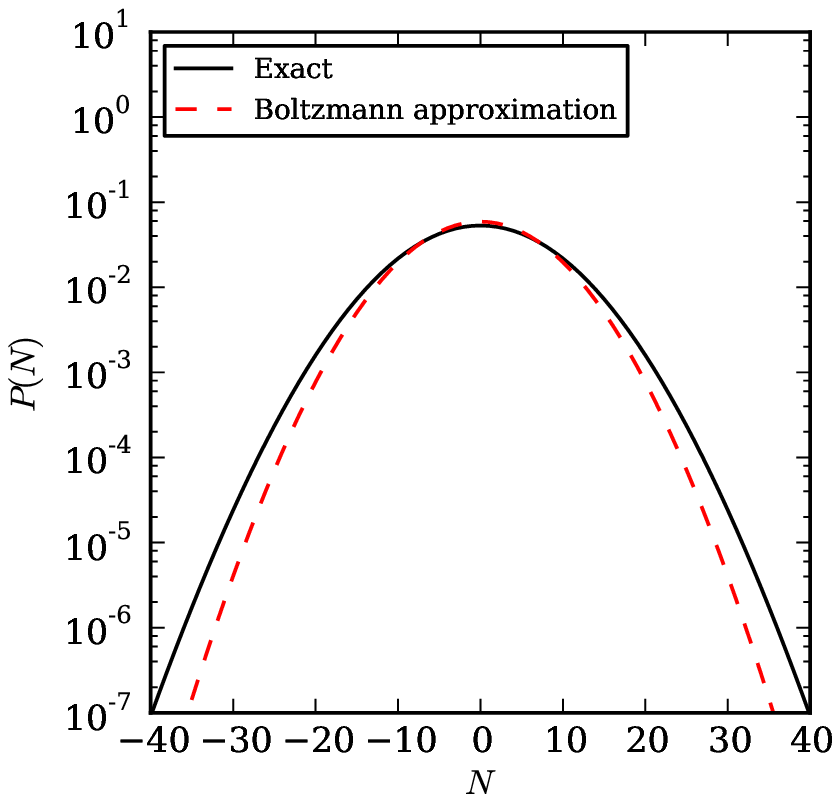}
\caption{
Left-hand figure: The ratios $R_5$ and $R_6$ from Eq. (\ref{Rratios})  in the HRG model along the freeze-out line.
Right-hand figure: The probability distributions for net-electric charge in a pion gas at
$T\simeq m_\pi$ calculated with quantum statistics (Exact) and in the
Boltzmann approximation.
}
\label{fig_rat}
\end{center}
\end{figure*}

 At
$\mu=0$, the  pion contribution to the $n$-th order cumulant of the  electric charge
reads
\begin{equation}
\chi^Q_n = \frac{m^2} {\pi^2 T^2} \sum_{l=1}^{\infty} l^{n-2} K_2 (l
\beta m_\pi).
\label{chipion}
\end{equation}
An approximation to $\chi^{Q}_{n}$ is obtained by truncating the
series (\ref{chipion}) at some order $l=l^*$. The
value of $l^*$, which yields a good approximation to the exact result, depends strongly on the
order of the cumulant.

At temperatures $T\simeq m_\pi$ the dominant
contribution to the cumulants $\chi^Q_n$ with $n\leq 5$ is due to the $l=1$
and $l=2$ terms.
In other words, in the HRG model, the next
to leading order Boltzmann approximation is sufficient to quantify
cumulants up to fifth order.
Consequently,  Eqs. (\ref{first}) and (\ref{last}) remain
valid if  the effective charge-2 pion contribution in the expansion (\ref{chipion}) is taken into
account. However,  this is not the case for the sixth- and higher-order cumulants.
The dominant contribution to $\chi^{Q}_{6}$ is due to the $l=3$ term in Eq. (\ref{chipion}). Consequently, Eq. (\ref{mg}) implies that
the sixth-order cumulant of the electric charge is well approximated by
\begin{eqnarray}
\chi_6^{Q} &\simeq&  \frac{1}{VT^3} \left[ \bar N_1+\bar N_{\bar 1} + 2^6( \bar N_2+\bar N_{2\pi}+\bar N_{\bar 2}+\bar N_{2\bar {\pi}}) 
+3^6(\bar N_{3\pi}+\bar N_{3\bar {\pi}}) \right],  
\label{mg6}
\end{eqnarray}
where $N_{1}$ ($N_{- 1}$) and $N_{2}$ ($N_{- 2}$) are the numbers of particles with charge $+1(-1)$ and   $+2(-2)$ respectively, while $\bar N_{k\pi}$ and  $\bar N_{k\bar {\pi}}$ are the $k$-th order corrections to the Boltzmann approximation for the charged pions numbers
\begin{equation}
\bar N_{k\pi}= \frac{VT^3}{2\pi^2}\left( {{m_\pi}\over {kT}}\right)^2 K_2\left({{km_\pi}\over T}\right)e^{k\beta \mu }.
\label{eq81}
\end{equation}
 The number of antipions,  $\bar N_{\bar {k\pi}}$, is obtained form Eq. (\ref{eq81}) by the replacement  $\mu\to -\mu$, where $\mu$ is the electric charge chemical potential.

The influence of quantum statistics on fluctuations of the electric charge
can be made more transparent  by considering the ratios:
\begin{equation}
R^Q_5 = \frac{\chi^Q_5 - \chi^Q_1}{\chi^Q_3 - \chi^Q_1}\,, ~~~~~~
R^Q_6 = \frac{\chi^Q_6 - \chi^Q_2}{\chi^Q_4 - \chi^Q_2}.
\label{Rratios}
\end{equation}
If  {\em only} doubly charged contributions are included in
 Eqs. (\ref{mg}--\ref{last}) and (\ref{mg6}),
the ratios $R^Q_5=R^Q_6=5$   are  independent of  temperature and chemical
potential. Any deviation from this value signal contributions emanating from the third- and higher-order corrections owing to
quantum statistics.

In Fig. \ref{fig_rat}-left, the  effect of quantum statistics  is illustrated
by the ratios $R_5$ and $R_6$ obtained in the HRG model along the chemical freeze-out line.
As shown, the ratio $R^Q_5$ is indeed within $<1$ \%  consistent with the result from Eqs.
(\ref{mg}--\ref{last}), while
$R^Q_6$ deviates by  up to  20 \% due to the missing effective triply charged pionic contribution to $\chi^{Q}_{6}$.

The effect of quantum statistics is also reflected in the net electric charge
probability distribution. In Fig. \ref{fig_rat}-right we show that the
inclusion of quantum statistics results in a broadening of
the net charge probability distribution of an ideal pion
gas, which is indeed very similar to the
effect of multi-charged particles.

In the HRG  the probability distribution
of the electric charge,   due to the relevant   contributions from the
second and the third order pionic  quantum statistics corrections,
can be explicitly  calculated  by applying   Eq. (\ref{p3}) instead of Eq. (\ref{p2}).


}


 \section{Conclusions}
We have analyzed the properties of the net charge probability
distributions in heavy-ion collisions within the hadron resonance gas
model. Within this model, these distributions and
the corresponding cumulants are linked directly to other observables,
and can be determined from the mean number of charged and multi-charged
particles.

Differences between the probability distributions for strangeness, electric charge and
net-baryon number were discussed and the
role of multi-charged particles and quantum statistics was explored. A
comparison of the UrQMD results for
the net baryon probability distribution in central Au-Au collisions at $\sqrt s_{NN}=$ 200 GeV
with data and with the hadron resonance gas model was also
presented. We find that the distribution obtained within the UrQMD model is much narrower than both
the experimental distribution and that of the hadron resonance gas.

The comparison of model results with data on  net proton distributions and
the  corresponding moments is particularly transparent. This is
because,  within the hadron resonance gas model, these quantities are
determined entirely  by the measured  yields of protons and
antiprotons. This offers an unambiguous route for confronting this
particular model with experimental data.


 We have presented model predictions for net proton probability
 distributions $P(N)$ in heavy ion collisions at RHIC and at LHC
 energies and computed the centrality dependence of $P(N)$ in Au-Au
 collisions at $\sqrt s_{NN}=$ 62.4 and 130 GeV. The results for
 $P(N)$ obtained in the hadron resonance gas along with the chemical
 freeze-out curve can be compared with  data.  Such a comparison would be an important
 test of the relation between the chemical freeze-out and the QCD
 cross-over line in heavy ion experiments.


\section*{ Acknowledgments}
We acknowledge discussions with Alexander Kalweit, Ilya Selyuzhenkov,
Johanna Stachel and Nu Xu. Comments from Chen Lizhu are also kindly
acknowledge.  K.R. acknowledges partial support of the Polish Ministry
of National Education (MEN).  The work of F.K. and V.S. was supported in part
by contract DE-AC02-98CH10886 with the U.S. Department of Energy.
B. F. acknowledges partial support by EMMI.


\end{document}